\documentclass[iop]{emulateapj}
\usepackage{graphicx}
\usepackage{amsmath}
\usepackage{atbegshi}
\def\hackaltaffiltext#1#2{\AtBeginShipoutNext{\footnotetext[#1]{#2}\stepcounter{footnote}}}

\begin{document}

\newcommand{\ocen}{$\omega$~Cen}

\title{Homogeneous analysis of globular clusters from the APOGEE
survey with the BACCHUS code $-$ III. \ocen}

\author{
Szabolcs~M{\'e}sz{\'a}ros\altaffilmark{1,2}, 
Thomas~Masseron\altaffilmark{3,4},
Jos{\'e}~G.~Fern{\'a}ndez-Trincado\altaffilmark{5,18}, 
D.~A.~Garc\'{\i}a-Hern{\'a}ndez\altaffilmark{3,4}, 
L{\'a}szl{\'o}~Szigeti\altaffilmark{1,2}, 
Katia~Cunha\altaffilmark{6,7}, 
Matthew~Shetrone\altaffilmark{8}, 
Verne~V.~Smith\altaffilmark{9}, 
Rachael~L.~Beaton\altaffilmark{10},
Timothy~C.~Beers\altaffilmark{11},
Joel~R.~Brownstein\altaffilmark{12},
Doug~Geisler\altaffilmark{13,14,15}, 
Christian~R.~Hayes\altaffilmark{16},
Henrik J\"onsson\altaffilmark{17},
Richard~R.~Lane\altaffilmark{18},
Steven~R.~Majewski\altaffilmark{19}, 
Dante~Minniti\altaffilmark{20,21,22}, 
Ricardo~R.~Munoz\altaffilmark{13},
Christian~Nitschelm\altaffilmark{23},
Alexandre~Roman-Lopes\altaffilmark{15},
Olga~Zamora\altaffilmark{3,4}
}
\altaffiltext{1}{ELTE E\"otv\"os Lor\'and University, Gothard Astrophysical Observatory, 9700 Szombathely, Szent Imre H. st. 112, Hungary}
\altaffiltext{2}{MTA-ELTE Exoplanet Research Group}
\altaffiltext{3}{Instituto de Astrof{\'{\i}}sica de Canarias (IAC), E-38205 La Laguna, Tenerife, Spain}
\altaffiltext{4}{Universidad de La Laguna (ULL), Departamento de Astrof\'{\i}sica, 38206 La Laguna, Tenerife, Spain}
\altaffiltext{5}{Instituto de Astronom\'ia y Ciencias Planetarias, Universidad de Atacama, Copayapu 485, Copiap\'o, Chile}
\altaffiltext{6}{Steward Observatory, University of Arizona, 933 North Cherry Avenue, Tucson, AZ 85721, USA} 
\altaffiltext{7}{Observat\'orio Nacional, S\~ao Crist\'ov\~ao, Rio de Janeiro, Brazil}
\altaffiltext{8}{University of Texas at Austin, McDonald Observatory, Fort Davis, TX 79734, USA}
\altaffiltext{9}{NOIRLab, Tucson, AZ 85719, USA}
\altaffiltext{10}{The Carnegie Observatories, 813 Santa Barbara Street, Pasadena, CA 91101, USA}
\altaffiltext{11}{Dept. of Physics and JINA Center for the Evolution of the Elements, University of Notre Dame, Notre Dame, IN 46556, USA}
\altaffiltext{12}{Dept. of Physics \& Astronomy, University of Utah, Salt Lake City, UT, 84112, USA}
\hackaltaffiltext{13}{Departamento de Astronom\'{\i}a, Universidad de Concepci\'on, Casilla 160-C, Concepci\'on, Chile}
\hackaltaffiltext{14}{Instituto de Investigación Multidisciplinario en Ciencia y Tecnolog\'{\i}a, Universidad de La Serena. Avenida Raúl Bitrán S/N, La Serena, Chile}
\hackaltaffiltext{15}{Departamento de Astronom\'{\i}a, Facultad de Ciencias, Universidad de La Serena. Av. Juan Cisternas 1200, La Serena, Chile}
\hackaltaffiltext{16}{Department of Astronomy, University of Washington, Box 351580, Seattle, WA 98195, USA}
\hackaltaffiltext{17}{Materials Science and Applied Mathematics, Malm\"o University, SE-205 06 Malm\"o, Sweden}
\hackaltaffiltext{18}{Centro de Investigaci{\'o}n en Astronom\'{\i}a, Universidad Bernardo O'Higgins, Avenida Viel 1497, Santiago, Chile}
\hackaltaffiltext{19}{Dept. of Astronomy, University of Virginia, Charlottesville, VA 22904-4325, USA}
\hackaltaffiltext{20}{Departamento de Ciencias Fisicas, Facultad de Ciencias Exactas, Universidad Andres Bello, Av. Fernandez Concha 700, Las Condes, Santiago, Chile}
\hackaltaffiltext{21}{Millennium Institute of Astrophysics, Av. Vicuna Mackenna 4860, 782-0436, Santiago, Chile}
\hackaltaffiltext{22}{Vatican Observatory, V00120 Vatican City State, Italy}
\hackaltaffiltext{23}{Centro de Astronom{\'i}a (CITEVA), Universidad de Antofagasta, Avenida Angamos 601, Antofagasta 1270300, Chile}

\begin{abstract}

We study the multiple populations of \ocen \ by using the abundances of Fe, C, N, O, Mg, Al, Si, K, Ca, and Ce from the 
high-resolution, high signal-to-noise (S/N$>$70) spectra of 982 red giant stars observed by the SDSS-IV/APOGEE-2 survey. We 
find that the shape of the Al-Mg and N-C anticorrelations changes as a function of metallicity, continuous for the metal-poor 
groups, but bimodal (or unimodal) at high metallicities. There are four Fe populations, similar to what has been found in 
previously published investigations, but we find seven 
populations based on Fe, Al, and Mg abundances. The evolution of Al in \ocen \ is compared to its evolution in the Milky Way 
and in five representative globular clusters. We find that the distribution of Al in metal-rich stars of \ocen \ 
closely follows what is observed in the Galaxy. Other $\alpha-$elements and C, N, O, and Ce are also compared to the 
Milky Way, and significantly elevated abundances are observed over what is found in the thick disk 
for almost all elements. However, we also find some stars
with high metallicity and low [Al/Fe], suggesting that \ocen \ could be the remnant core of a dwarf galaxy,
but the existence of these peculiar stars needs an independent confirmation. We also confirm the 
increase in the sum of CNO as a function of metallicity previously reported in the literature and find that the [C/N] ratio 
appears to show opposite correlations between Al-poor and Al-rich stars as a function of metallicity.

\end{abstract}

\section{Introduction}

Globular clusters (GCs) are long known to host multiple populations of stars that have different chemical compositions. 
In the last decade, these multiple populations (MPs) have been well explored by large spectroscopic surveys 
\citep[e.g.,][]{carretta02, carretta03, carretta01, meszaros03, 
pancino01, masseron01, meszaros05, horta01} and photometric  \citep[e.g.,][]{piotto01, sara01, piotto02, milone02, soto01}.
Recently, \citet{bastian03} provided an overview on MPs in GCs. 

Even though the multiple populations in GCs usually have Na, O, Al, Mg, C and N abundance variations, there are very few 
globular clusters known to have a variance in Fe larger than the abundance measurement errors. One such cluster is \ocen \;
having stars with a wide range in metallicity ($-$2.2 $<$ [Fe/H] $< -$0.6) this cluster has been studied in detail by several 
groups who identified four main populations with different metallicities
(\citet{norris02, pancino04, johnson02, marino03}).
The existence of several populations in \ocen \ has led to the conclusion that this cluster was probably massive enough, at 
some time, to retain the ejecta of supernovae, which would have progressively enriched each new star generation with the Fe 
produced at the end of each burst of star formation. 

The number of populations that exist in \ocen \ remains an open question and strongly depends on the selected element or 
families of elements used to identify these groups. 
For example, \citet{gratton04} identified six populations that belong to three main groups in \ocen \ by using the $k-$means 
algorithm and the abundances from \citet{johnson02}. 
\ocen \ exhibits large star-to-star variations in several light elements, including C, N, O, Na, Mg, Al, and Si and also 
s-process elements \citep{norris02, smith00, johnson02, stanford01, marino01}. 
While this is similar to other regular GCs, \ocen \ is unique in the sense that each group in metallicity shows its own 
Na-O anticorrelation; the more metal-rich populations have a higher percentage of Na-rich and O-poor stars than metal-poor 
ones, resulting in a slightly more extended Na-O anticorrelation than the one observed in the metal-poor 
regime \citep{marino03}. 
In addition, a correlation between the sum of CNO and s-process elements with metallicity has also been observed for 
\ocen \ \citep{johnson02, marino03}. 
The split main sequence in \ocen \ is suspected to be accompanied by a large variation in the He content, based 
on fitting stellar models on Hubble Space Telescope observations \citep{bellini01, king01}. 
Helium enrichment also generally correlates with [Na/Fe] and [Al/Fe] \citep{dupree01}, with classic first population stars 
having Y$<$0.22, in contrast to those stars that are enriched in Na, for which \citep{dupree01} found Y=0.39$-$0.44. 

Kiskunfelegyhaza

\begin{deluxetable*}{llrrrrrrrrrrr}
\tabletypesize{\scriptsize}
\tablewidth{0pt}
\tablecaption{Atmospheric Parameters and Abundances of Individual Stars}
\tablehead{
\colhead{2MASS ID} & \colhead{Cluster} & 
\colhead{Status} & \colhead{T$_{\rm eff}$} &
\colhead{log g} & \colhead{[Fe/H]} & 
\colhead{$\sigma_{\rm [Fe/H]}$} & \colhead{[C/Fe]} & 
\colhead{limit\tablenotemark{a}} & \colhead{$\sigma_{\rm [C/Fe]}$} & 
\colhead{$N_{\rm C}$} & \colhead{[N/Fe]} & \colhead{...} \\
\colhead{} & \colhead{} & 
\colhead{} & \colhead{(K)} &
\colhead{(cgs)} & \colhead{(dex)} & 
\colhead{(dex)} & \colhead{(dex)} & 
\colhead{} & \colhead{(dex)} & 
\colhead{(dex)} & \colhead{(dex)} & \colhead{...}
}
\startdata
2M13242154-4738394 	&  Omegacen	& RGB & 4657 &   1.44 &  $-$1.595      &   0.04   &     \nodata & 	0   &     \nodata   &   0   &     \nodata  & \\
2M13242622-4729383 	&  Omegacen	& RGB & 5594 &   3.51 &  $-$1.214      &   0.12   &     \nodata & 	0   &     \nodata   &   0   &     \nodata  &  \\
2M13243074-4724264 	&  Omegacen	& RGB & 3993 &   0.17 &   $-$1.65      &  0.067   &   	$-$0.17 & 	1   &   0.046   	&   4   &   0.625  & \\
2M13243844-4736586 	&  Omegacen	& RGB & 4978 &   2.09 &  $-$1.783      &  0.274   &   	0.853 & 	1   &   0.049  	 	&   1   &     \nodata  & \\
2M13244450-4735071 	&  Omegacen	& RGB & 5271 &   2.79 &  $-$1.706      &  0.071   &     \nodata & 	0   &     \nodata   &   0   &     \nodata  &
\enddata
\tablecomments{This table is available in its entirety in machine-readable form in the online journal. A portion 
is shown here, with reduced number of columns, for guidance regarding its form and content. 
Star identification from \citet{carretta03} was added in the last column.}
\tablenotetext{a}{The number of lines used in the abundances analysis from BACCHUS \citep{masseron02}.}
\end{deluxetable*}

\begin{deluxetable}{lcc}[!ht]
\tabletypesize{\scriptsize}
\tablewidth{0pt}
\tablecaption{Abundance Averages and Scatter}
\tablehead{
\colhead{Parameter} & \colhead{\citet{meszaros05}} & 
\colhead{This paper} 
}
\startdata
N$_{\rm 1}$\tablenotemark{a}	& 898 		& 1141	\\
N$_{\rm 2}$\tablenotemark{b}	& 775 		& 982	\\
${\rm [Fe/H]}$ average					& $-$1.511 	& $-$1.528	\\
${\rm [Fe/H]}$ scatter 					& 0.205 	& 0.233	\\
${\rm [Fe/H]}$ error\tablenotemark{c}	& 0.077 	& 0.079	\\
${\rm [Al/Fe]}$ average	                & 0.586 	& 0.497	\\
${\rm [Al/Fe]}$ scatter                 & 0.533 	& 0.566	\\
${\rm [Al/Fe]}$ average	$>$0.3dex       & 0.935 	& 0.927	\\
${\rm [Al/Fe]}$ scatter $>$0.3dex	    & 0.389 	& 0.442	\\
${\rm [Al/Fe]}$ average $<$0.3dex		& 0.058 	& 0.020	\\
f$_{\rm enriched}$\tablenotemark{d}     & 0.603 	& 0.527	\\
${\rm [(Mg+Al+SI)/Fe]}$ average         & 0.413 	& 0.410	\\
${\rm [(Mg+Al+SI)/Fe]}$ scatter			& 0.096 	& 0.098	\\
${\rm [N/Fe]}$ average                  & 1.273 	& 1.334	\\
${\rm [N/Fe]}$ scatter                  & 0.452 	& 0.520	\\
${\rm [(C+N+O)/Fe]}$ average            & 0.642 	& 0.699	\\
${\rm [(C+N+O)/Fe]}$ scatter			& 0.177 	& 0.216	\\
\enddata
\tablecomments{This table lists statistics for \ocen \ from \citet{meszaros05} compared to this paper.}
\tablenotetext{a}{The number of all stars in our sample.}
\tablenotetext{b}{The number of stars with S/N$>$70.}
\tablenotetext{c}{The average uncertainty of [Fe/H].}
\tablenotetext{d}{f$_{\rm enriched}$ = N$_{\rm SG}$/N$_{\rm tot}$ }
\end{deluxetable}

Considering all the chemical signatures and observational evidences discussed it is apparent that the formation of \ocen \ was probably more complex than any other Galactic GC. 
It is possible that \ocen \ was accreted by the Milky Way but the question remains if \ocen \  also interacted with the Milky Way. 
There are two leading scenarios proposed for the formation of \ocen; the first is that it is the remnant core of a 
dwarf galaxy \citep{bekki01}, and the second a merger of multiple clusters \citep{bergh01, lee01}. As a combination of 
the two, modeling of the merging scenario of clusters in a dwarf galaxy has shown great promise lately, as it has been 
able to reproduce the observed properties of \ocen \ \citep{mast01}. 

The more metal rich stars in \ocen \ are concentrated in the southern part of the cluster. 
\citet{pancino02, pancino03} found that the three main stellar sub-populations of \ocen \ have different spatial 
distributions making it one of the very few clusters currently known in which metal-rich stars have a more extended 
spatial distribution than metal-poor stars. This spatial coverage was studied in detail by \citet{calamida01}, who 
concluded that merging scenario is a viable explanation for the formation of \ocen. Evidence of interaction with the Milky 
Way was eventually discovered by \citet{ibata01} who proved with N-body simulations that the ‘Fimbulthul’ structure, 
a stream of 309 stars that surround the inner Galaxy \citep{ibata02}, is the tidal stream of \ocen. 

In this paper, we will revisit the metallicity distribution of \ocen \ in Section~3, examine the shape of the Al-Mg, 
the N-C anti-correlations, and deep mixing as a function of metallicity in Sections~4 and 7. 
Our goal is to investigate the formation of \ocen \ by deriving abundances 
of its multiple populations. We aim to do this by comparing the [Al/Fe] distribution in \ocen \ and the Milky Way. 
These discussions can be found in Sections~5 and 6.

\section{The \ocen \ Sample}

This study is a continuation of our recent work \citep{meszaros05} analyzing APOGEE data in \ocen \. 
Similarly to our previous studies, we select stars based on their 
radial velocity and their distance from the cluster center, but we did not limit the metallicity of the 
selected stars for \ocen. In radial velocity, we 
required stars to be within three times the velocity dispersion of the mean cluster velocity, and in distance we 
required stars to be within the tidal radius. We also used the average cluster radial velocity and its scatter 
from Gaia DR2 \citep{baum01} rather than from \citet{harris01}. In addition, we introduced a fourth step that is based 
on selecting stars that have proper motion within a 2.0 mas yr$^{-1}$ range 
around the cluster average proper motion from the Gaia DR2 catalog \citep{gaia01}. 

In APOGEE DR16 \citep{ahumada01, hol03} there were 243 additional targets belonging to \ocen \ that had not been included in 
\citet{meszaros05}; these stars are now analyzed here. In \citet{meszaros05} we only gave a short overview 
of \ocen, without separately discussing its populations. The majority of these stars belong to the most 
metal-poor and most metal-rich populations, making it possible to examine the chemical properties of these populations 
in more detail than otherwise would have been possible. We derived their atmospheric parameters and used the BACCHUS 
\citet{masseron02} pipeline to obtain their chemical abundances. The exact same setup from \citet{masseron01} and 
\citet{meszaros05} was used, such that the results from these additional targets  (207 of them with S/N$>$70) are on 
the same scale and can be merged with the previous results from \citet{meszaros05} in order to further examine the 
statistics of multiple populations in \ocen. Similarly to \citet{masseron01} and 
\citet{meszaros05} we used \citet{asplund01} for our Solar abundance reference table. 
The complete set of stellar parameters and abundances can be found in 
Table~1, which includes stars from \citet{meszaros05}; there are 1141 stars, 982 of them have S/N$>$70, 
compared to 898 stars and 775 with S/N$>$70 published in \citet{meszaros05}. 
We limit our discussion to stars with S/N$>$70, and also apply the same quality assurance cuts in atmospheric 
parameters as in our previous papers.

\begin{figure}
\centering
\includegraphics[width=2.45in,angle=270]{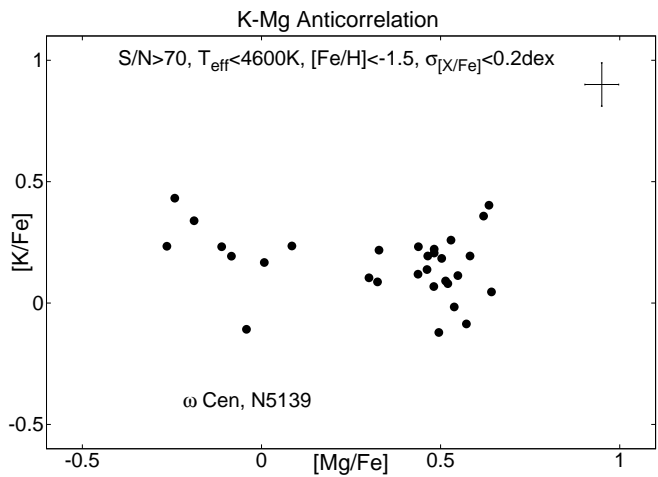}
\caption{The observed K-Mg anticorrelation in \ocen. 
}
\label{fig:kmg}
\end{figure}

\begin{figure*}
\centering
\includegraphics[width=4.4in,angle=270]{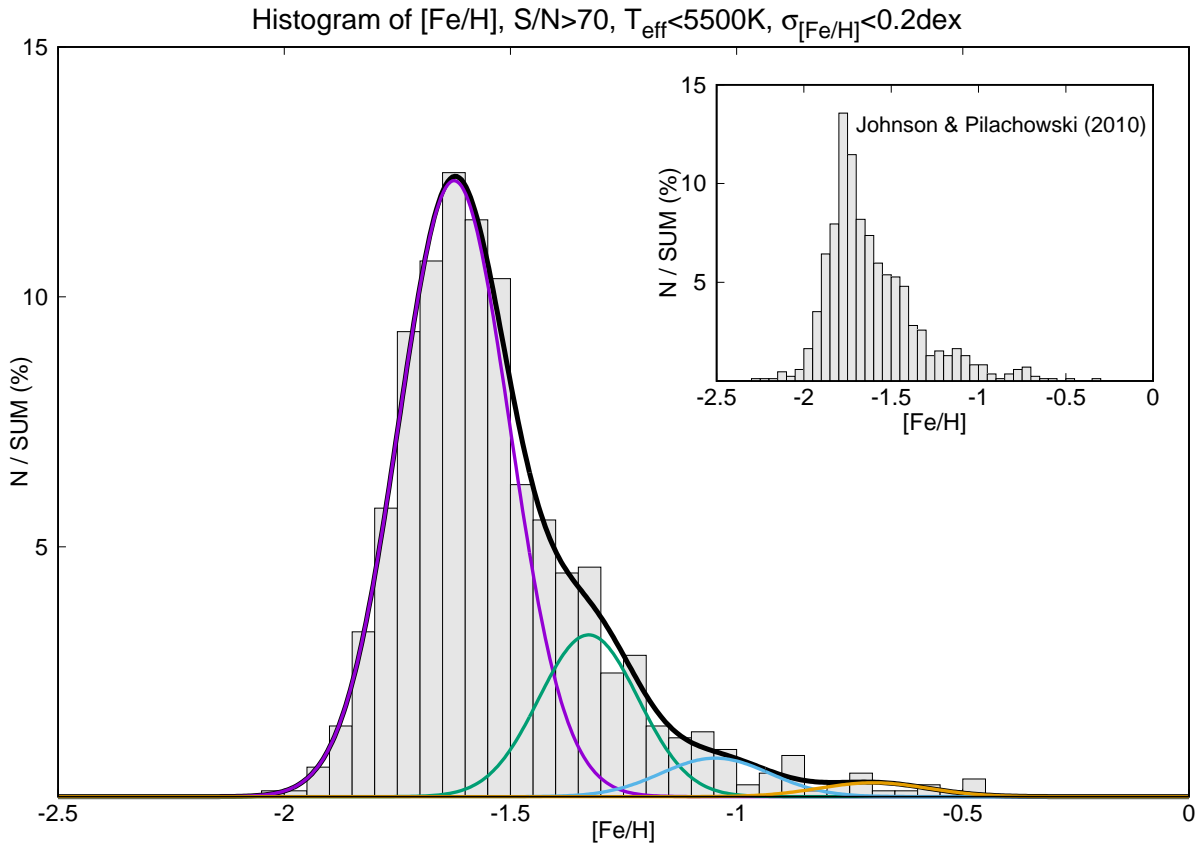}
\caption{The metallicity distribution of \ocen. The overall structure can be best fit with four populations with different 
Fe abundances. Our measurements show a very similar histogram to that of \citet{johnson02} seen in the inset. 
}
\label{fig:met}
\end{figure*}

\begin{figure*}
\centering
\includegraphics[width=4.4in,angle=270]{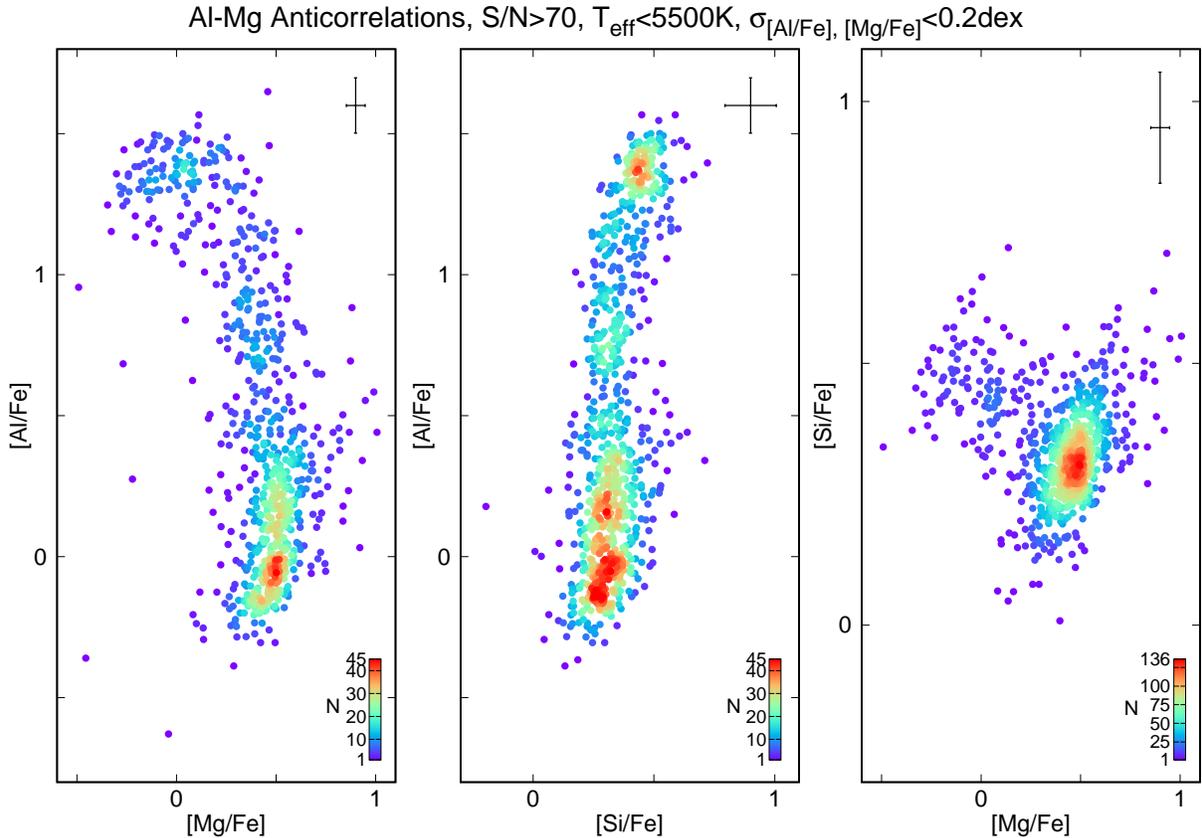}
\caption{The effect of the Mg-Al cycle in \ocen \ color coded by the number of stars in a 
$\pm$0.05~dex range around each star. The turnaround of Al abundances at very low Mg abundance values is 
clearly visible along with density peaks marking various star populations related to star forming episodes. The 
average errors are plotted in the top right corner of each panel.
}
\label{fig:almg}
\end{figure*}

\begin{figure*}
\centering
\includegraphics[width=4.4in,angle=270]{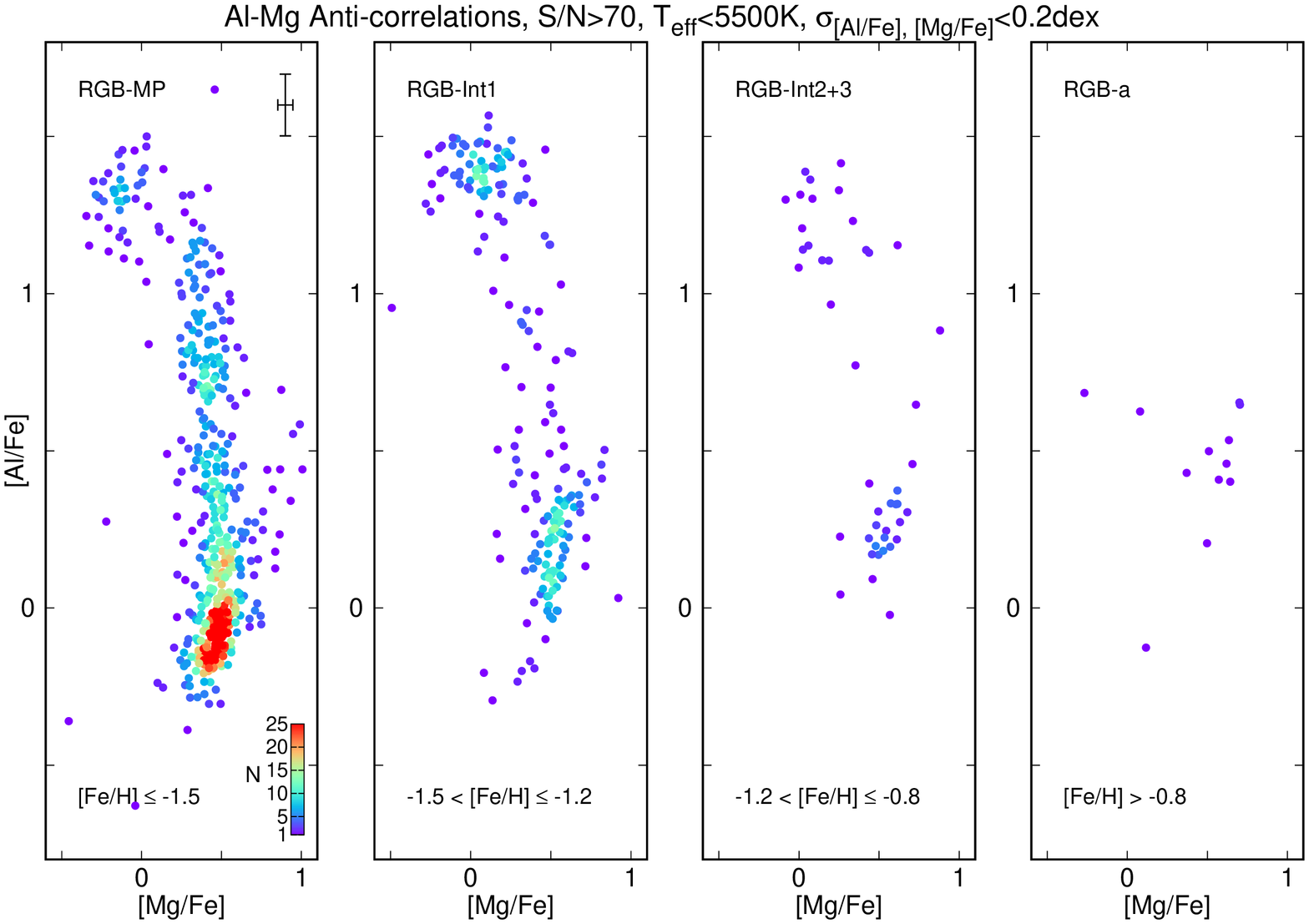}
\caption{The Mg-Al anticorrelation for each population defined in the metallicity histogram. The shape of the anticorrelation
changes with the metallicity, it is nearly continuous when [Fe/H]$<-$1.2, then becomes bimodal between 
$-$1.2$<$[Fe/H]$<-$0.8 and nearly constant above [Fe/H]$>-$0.8.
}
\label{fig:alfe}
\end{figure*}

A comparison of the average Fe, Al, Mg+Al+Si, N, and C+N+O abundances obtained for entire sample of \ocen \ stars with the 
previous results for the smaller sample in \citet{meszaros05} is presented in Table~2. 
Overall the changes in the abundance averages resulting from the addition of the studied 243 targets to the  
\citet{meszaros05} sample do not change the averages beyond the uncertinaties and do not affect the science 
results of \citet{meszaros05}.
We note, however, that since most of the additional stars belong to the metal-poor or metal-rich population, 
the scatter of [Fe/H] increased slightly from 0.205 to 0.233~dex; the Al/Fe average abundances decreased by 
0.089~dex because the majority of the added stars have low Al abundances, which also leads to a decrease of 
f$_{\rm enriched}$ (N$_{\rm SG}$/N$_{\rm tot}$) from 0.603 to 0.527. 
Such changes are expected to appear with more stars added to the sample with new observations, as statistics 
evolve and become more complete when the sample size increases.

\subsection{The K-Mg anticorrelation}

A possible weak anti-correlation between K and Mg in \ocen \ was reported in \citet{meszaros05}. However, 
this possibility is weakened by the fact that the two K I lines found in the APOGEE region are often blended 
and also fairly weak at high temperatures and low metallicities. With the addition of a few more stars, the 
resulting anticorrelation in this study (seen in Figure~\ref{fig:kmg}) is quite similar to what was published 
in \citet{meszaros05}; while there are now more stars observed with [Mg/Fe]$<$0.1, the anticorrelation 
signature does not seem to be clearer than in that in \citet{meszaros05} as we do not find new stars with 
high K and low Mg and we still believe that the K enhancement of the Mg-poor stars in \ocen \ cannot be 
convincingly claimed from our measurements.

\section{The metallicity distribution}

As mentioned in the introduction, many spectroscopic and photometric studies in the literature have demonstrated 
that  \ocen \ has a wide spread in metallicity. In particular, the study by \citet{johnson02} presented [Fe/H] 
measurements for a nearly complete sample of 867 \ocen \ giants with V$<$13.5, covering the cluster's full 
metallicity range. Our sample also covers almost all stars with V$<$13.5, but it includes a larger number of stars with V$<$16.
Their sample consisted of 867 giants, while ours is slightly larger at 982 giants
In Figure~\ref{fig:met} we show the metallicity distribution obtained in this study and as a comparison we 
also show on the top part of the figure the metallicity distribution from \citet{johnson02}. 
To compare our sample with that of \citet{johnson02} we used the same bin size of 0.05~dex. 
The general trend in the two metallicity distributions (both obtained from high-resolution spectra) agree quite 
well: overall the two distributions have a main peak roughly at the same metallicity and a tail that extends to 
metallicities around -0.5. \citet{johnson02} identified five separate populations based on peaks in their 
metallicity distribution which are located at: [Fe/H]$=-$1.75, $-$1.50, $-$1.15, $-$1.05, and $-$0.75 (the 
peaks at $-$1.15 and $-$1.05 were combined because of the difficulty in de-blending these two populations). 
Our metallicity distribution is very similar to theirs and four populations can also be identified, which 
correspond to the following peaks in metallicity: $-$1.65, $-1.35$, $-1.05$, and $-$0.7~dex. The metallicity 
peaks in our study are systematically higher by about 0.1~dex than the ones from \citet{johnson02}. Such an 
offset can be considered as quite small, given the very different sets of lines and methodologies used for 
deriving metallicities from the H band (APOGEE) and the optical spectra \citep{meszaros05}. 
Following \citet{sollima01} and accounting for the 0.1 systematic offset in metallicity, these four populations 
are named: RGB-MP ([Fe/H]$<=-$1.5), RGB-Int1 ($-$1.5$<$[Fe/H]$<=-$1.2), RGB-Int2+3 ($-$1.2$<$[Fe/H]$<=-$0.8), 
and RGB-a ([Fe/H]$>-$0.8).


\section{Multiple Populations Based on Al and Mg Abundances}

The result of the Mg-Al cycle is observed in almost all GCs that are more metal-poor than [Fe/H]$<-$1
\citep{shetrone01, carretta02, meszaros03, meszaros05}. 

According to \citet{ventura04} an anti-correlation between the 
abundances of Mg-Al happens because the Mg-Al cycle in high mass AGB stars requires such high temperatures 
($>$70 million Kelvin) to operate that this can only be reached in the core of low metallicity stellar polluters 
\citep{ventura04}. While this AGB scenario explains the shape of the Mg-Al anticorrelation relatively well it 
does not explain other observed features, for example the Na-O anticorrelation. In fact, none of the currently 
available models can fully explain all the observed chemical properties of MPs in GCs \citep{bastian03}. 
In this paper, we restrict our discussion to the AGB scenario, as most of our discussion lies on the observed 
properties of Mg and Al abundances of each population. We note, however, 
that an Al-Mg anti-correlation has been reported for the more metal-rich Bulge GC NGC~6553 \citep{tang01, schiavon01}, 
and that some atypical Al-rich field stars with [Fe/H]$>-$1 (although with Mg abundances much lower than GCs of 
similar metallicity) have been discovered in our Galaxy \citep{trin02, schiavon01, horta01}.

While the shape of the Al-Mg anticorrelation exhibits significant diversity from cluster to cluster, (see 
discussion \citet{meszaros05}), no globular cluster has an Al-Mg anti-correlation as complex as \ocen. The 
Al-Mg abundances for the \ocen \ sample are shown in Figure~\ref{fig:almg} (left panel), along with the observed 
Al-Si correlation (middle panel) and the Si-Mg anti-correlation (right panel). The general shape of these trends 
is very similar to what was presented in our previous study \citep{meszaros05}, even though we have roughly 25\% 
more stars than previously. 
\citet{masseron01} discovered that some stars in M15 and M92 show an extreme Mg depletion with some Si enhancement, 
while at the same time being Al depleted relative to the most Al rich stars in these clusters, displaying a turnover 
in the Al-Mg diagram. Later, \citet{meszaros05} found the same effect for \ocen, using a smaller sample size than 
the one used here. 
We explained the presence of this turnaround as the partial depletion of Al in their progenitors by very hot proton-capture 
nucleosynthetic processes occurring above 80~MK temperatures. 

\subsection{Metallicity Dependence}

The complex nature of the Al abundance as a function of metallicity in \ocen \ stars has been previously discussed 
in the literature, for an overview see \citet{johnson02}. In this study we derive [Al/Fe] abundances for 873 giant 
stars (having APOGEE spectra with S/N$>$70), allowing us to map the [Al/Fe] abundance distribution in more detail 
than previous studies (of which, the largest to date is \citet{johnson02}, who reported Al abundances for 332 stars in \ocen). 

In Section~3, following the method used by \citet{johnson02}, 
we defined four populations for \ocen \ based on the peaks found in the Fe abundance distribution plotted in 
Figure~\ref{fig:met}. In Figure~\ref{fig:alfe} we now divide the Al-Mg results into these four metallicity 
groups. Overall, the structure of the Al abundance distribution in this study is very similar to that of \citet{johnson02}.
The turnover of [Al/Fe] at low values fo [Mg/Fe] is observed only in the most metal-poor group, RGB-MP, below 
[Fe/H]$<-$1.5, again confirming that the very hot Mg-Al burning only occurred in metal-poor AGB stars. 
The extended sample (see Section~2) adds more stars to this most metal-poor 
population and confirms the existence of the turnaround of Al abundances in \ocen.

It is apparent that the shape 
of the Al-Mg anti-correlation changes with metallicity. RGB-MP shows a nearly 
continuous distribution, while RGB-Int1 becomes more bimodal than continuous and this trend continues to 
RGB-Int2+3, which is clearly bimodal. The most metal-rich group, RGB-a, shows no significant Al abundance spread. 
This is similar to what is seen in other metal-rich galactic GCs, with metallicities above [Fe/H]$=-$0.8, 
where the Mg-Al burning might not happen, or, its effects may not visible in the logarithmic abundance scale at 
such high metallicities \citep{ventura04, meszaros05}. 

To put the \ocen \ results discussed above in context we note that bimodal clusters are observed at various 
metallicities  \citep{carretta02, meszaros03, meszaros05}, but if we consider the 31 clusters in \citet{meszaros05}, 
no clear correlation is found between metallicity and the shape of the Al-Mg anti-correlation for clusters with 
[Fe/H]$<-$0.9. For example, M53 ([Fe/H]=$-$1.888) is clearly bimodal, while M5 ([Fe/H]=$-$1.178) appears continuous. 
In \ocen, the Al-Mg anti-correlation apparently becomes more and more bimodal as the metallicity increases. This 
will be further discussed in Section~5.

\subsection{Identifying Multiple Populations}

\begin{figure}
\centering
\includegraphics[width=2.45in,angle=270]{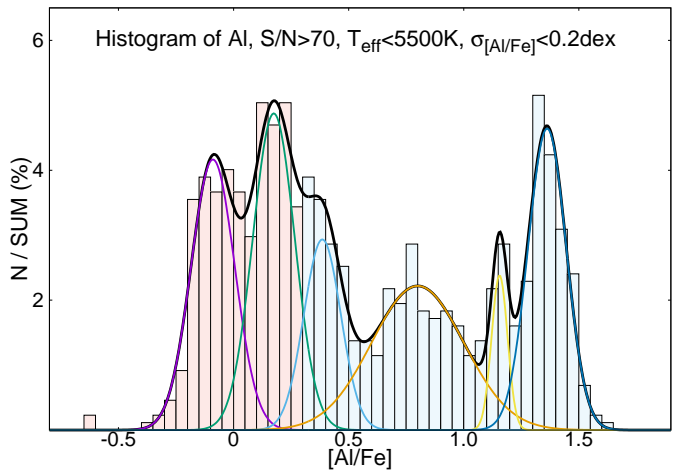}
\caption{Left panel: The histogram of Al in 0.05~bins in \ocen. There are six populations identified by fitting Gaussian functions 
to the Al histogram. These populations are marked on the Al-Si correlation in the right panel, the radius of the circles is 
equal to the FWHM of the fitted functions.
}
\label{fig:alhist}
\end{figure}

\begin{figure*}
\centering
\includegraphics[width=4.4in,angle=270]{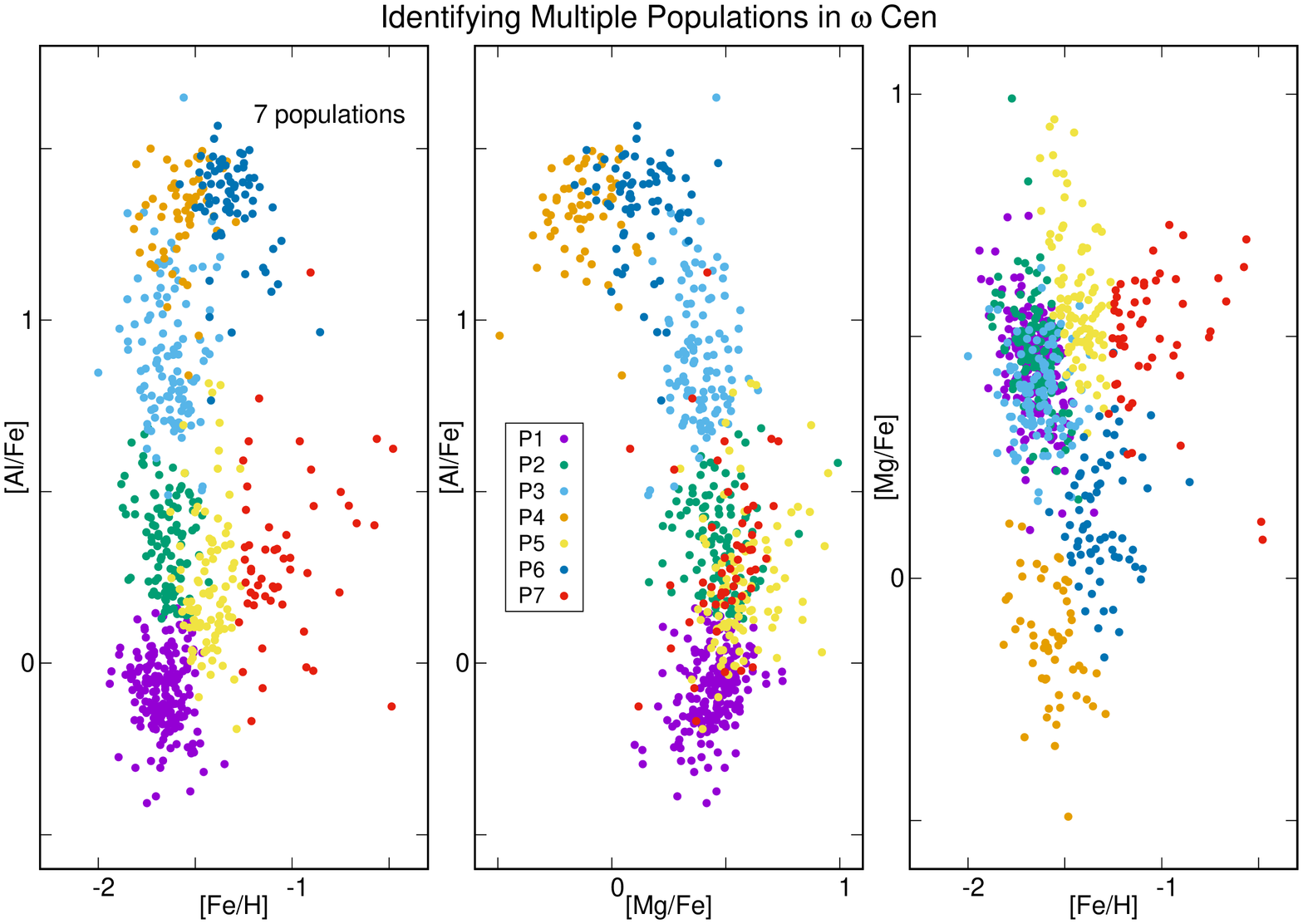}
\caption{The distribution of stars in \ocen \ from seven populations that separates stars in the Fe-Al-Mg 
abundance space.
}
\label{fig:alpop}
\end{figure*}

While the increased number of analyzed stars do not affect the general shape of the Al-Mg anti-correlation, 
it allows us to analyze the Al histogram in more detail than done by \citet{meszaros05}. Both the Al-Mg and Al-Si 
relations have continuous distribution of Al with well defined density peaks clearly visible in Figure~\ref{fig:almg}, 
but the Al histogram by itself is not enough to give a more complete picture of the distribution of Al, as it 
integrates stars with different [Mg/Fe] and [Fe/H] together. This can be seen in Figure~\ref{fig:alhist}. 

In \citet{meszaros05} three main populations in \ocen \ could be distinguished based on their Al abundance distributions, 
these can now be refined using the extended APOGEE data set. The three populations found previously peaked at around 
[Al/Fe]=0.1, 0.75 and 1.4, but now an examination of, for example, the [Al/Fe]-[Si/Fe] correlation (middle panel of 
Figure~\ref{fig:almg}), reveals that the main group centered around [Al/Fe]=0.1 can be split further into two subgroups. 
This in contrast with the majority of GCs. In \citet{meszaros05} an initial division of FG and SG stars was done 
with a simple cut at [Al/Fe]=0.3~dex; such division in [Al/Fe] works well for most GCs in which the effect of Al 
pollution can be observed. The FG stars with [Al/Fe]$<$0.3 could not be divided any further in any of these 
clusters, within the uncertainties \citep{meszaros05}. However, this is not the case for \ocen \ .

It is not as straightforward to simply define FG and SG stars in \ocen \ as we may be seeing at least 
six different populations based on the Al histogram. In Figure~\ref{fig:alhist} six Gaussian functions were fitted to the peaks 
in [Al/Fe] in the histogram. The first group, originally defined around the peak at [Al/Fe]=0.1 in \citet{meszaros05}, 
is the most numerous and can be divided into three further groups with peaks at [Al/Fe]=$-$0.09, 0.17, and 0.39. The 
second group, originally at [Al/Fe]=0.75, has the widest distribution in Al abundances, but cannot be further split 
into subgroups using the Al abundances alone; it peaks at [Al/Fe]=0.8 in Figure~\ref{fig:alhist}. The new APOGEE 
data more clearly indicate the presence of two peaks in the third group (originally centered around [Al/Fe]=1.4 
\citet{meszaros05}): one that peaks 
at [Al/Fe]=1.16, and another one at [Al/Fe]=1.36. We note, however, that the latter subgroup is well defined, 
while the former contains only a few stars.

\begin{deluxetable}{lcccccc}[!ht]
\tabletypesize{\scriptsize}
\tablewidth{0pt}
\tablecaption{Abundance Averages of Populations}
\tablehead{
\colhead{Pop.} & N\tablenotemark{a} & [Fe/H] & [Mg/Fe] & [Al/Fe]  \\
& & average & average & average 
}
\startdata
P1 & 190 & $-$1.663 & 0.445 & $-$0.080 \\
P2 & 107 & $-$1.651 & 0.470 & 0.352  	\\
P3 & 107 & $-$1.634 & 0.388 & 0.898  	\\
P4 & 56  & $-$1.572 & $-$0.128 & 1.307 \\
P5 & 95  & $-$1.431 & 0.578 & 0.229  \\
P6 & 68  & $-$1.312 & 0.124 & 1.334  \\
P7 & 52  & $-$1.040 & 0.500 & 0.307  	\\
\enddata
\tablecomments{This table lists the average [Fe/H], [Mg/Fe] and [Al/Fe] of the multiple populations in \ocen.}
\tablenotetext{a}{The number of stars in each population.}
\end{deluxetable}

Similarly to \citet{gratton04} we carried out a population analysis using the popular 
K-means clustering algorithm implemented in R \citep{Steinhaus56}. We investigated the elements Al, Mg and Fe, 
and also tested the addition of Si, which has no effect on the number of populations found. The K-means clustering 
algorithm revealed the existence of seven populations, or, one more population than found from the analysis of the 
[Al/Fe] histogram. Figure~\ref{fig:alpop} shows the distribution of the seven populations in the Fe-Al-Mg space. 
The difference between the Gaussian peak analysis versus the K-means analysis is limited to the groups with peaks 
at [Al/Fe]=0.17, and 0.39 in the [Al/Fe] histogram. These groups split into three sub-groups after the K-means 
analysis was able to distinguish between groups with different metallicities, resulting in an increase from six 
to seven populations. The abundance averages and adopted names for the seven populations are found in Table~3.

By comparing results in this study with those from \citet{gratton04} we conclude that the same seven populations 
found here can be divided into three groups based on their metallicities. This reinforces the findings of the two 
independent studies, which are based on different sets of elements, \citet{gratton04} used [Na/O] and [La/H], while 
this study [Fe/H], [Mg/Fe], and [Al/Fe]. Populations P1, P2, P3 and P4 belong to the most metal-poor group of stars 
(Table~3), RGB-MP from \citet{sollima01}, P5 and P6 have intermediate metallicities, these belong to the RGB-Int1 
and RGB-Int2+3 groups, and P7 is the most metal-rich (RGB-a) that is not found to further divided into sub-populations 
based on our data.

From Figure~\ref{fig:alpop} we can identify which population is responsible for the turnaround in the Al-Mg 
anti-correlation discussed in Section~4: the group of stars belonging to the 4th population in our sample, P4, 
which is shown as orange dots in Figure~\ref{fig:alpop}. Most of stars in P4 belong to the metal-poor population 
RGB-MP, with some stars from RGB-Int1.

\section{Multiple Populations Based on N and C}

\begin{figure}[!ht]
\centering
\includegraphics[width=2.45in,angle=270]{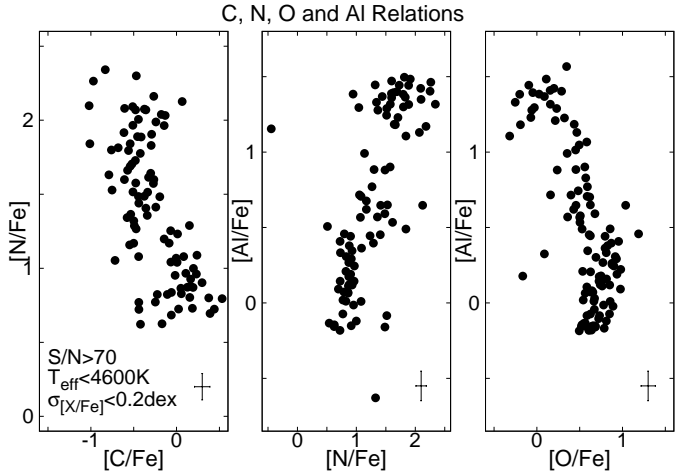}
\caption{The relationship between C, N, O, and Al in \ocen. As expected N is anti-correlated with C, while Al is 
correlated with N, but anti-correlated with O.
}
\label{fig:nc}
\end{figure}

\begin{figure}[!ht]
\centering
\includegraphics[width=2.45in,angle=270]{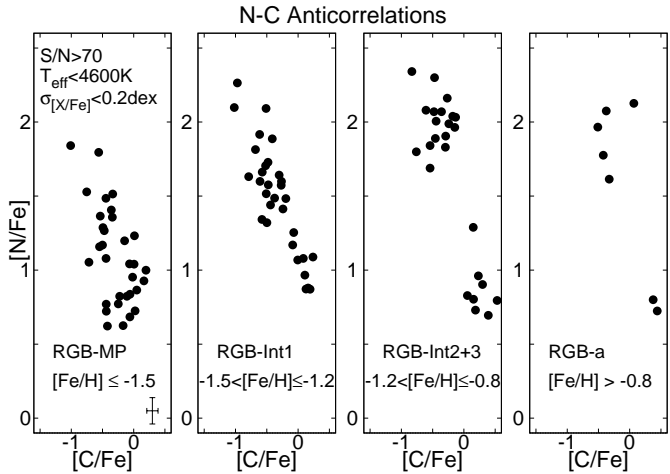}
\caption{The N-C anti-correlation in the four populations defined from the Fe histogram. The anticorrelation is 
nearly continuous when [Fe/H]$<$-0.8, then becomes bimodal above [Fe/H]$>-$0.8.
}
\label{fig:ncfe}
\end{figure}

In addition to the Al-Mg anti-correlation, C-N-O-Al abundance relations can also be studied using APOGEE spectra. 
The N-C and Al-O anti-correlations and the Al-N correlation obtained for \ocen \ are shown in Figure~\ref{fig:nc} and 
Figure~\ref{fig:ncfe}. These are quite similar to those in \citet{meszaros05} despite the increased star count. 
The N-C anti-correlation, however, is much clearer in our APOGEE data set than previously observed by 
\citet{marino01}. 

Similarly to the Al-Mg anti-correlation, the N-C anti-correlation also depends on the metallicity. However, 
in this study it is not possible to probe the N-C anti-correlation in the most metal-poor groups because the 
molecular lines of CN, OH and CO disappear in the APOGEE spectra of those stars having metallicities below 
approximately [Fe/H]$<-$1.8. The distribution of N abundances are quite continuous for [Fe/H]$<-$1.2, but 
becomes bimodal for the most metal-rich stars with [Fe/H]$>-$1.2. This bimodality occurs in both the N-C 
and Al-Mg planes and, in both, not at a significantly different metallicity value.
There is a difference in the most metal-rich group, RGB-a, which has no significant Al spread beyond our 
uncertainties, but stars with high N are clearly observed. It is worth noting that some of the stars with 
high N in the RGB-a population have 
$\sigma_[Fe/H]>$ 0.2dex, due to difficulties in fitting molecular lines below T$_{\rm eff}<$4200~K.

\subsection{Deep Mixing}

\begin{figure}
\centering
\includegraphics[width=3.45in,angle=0]{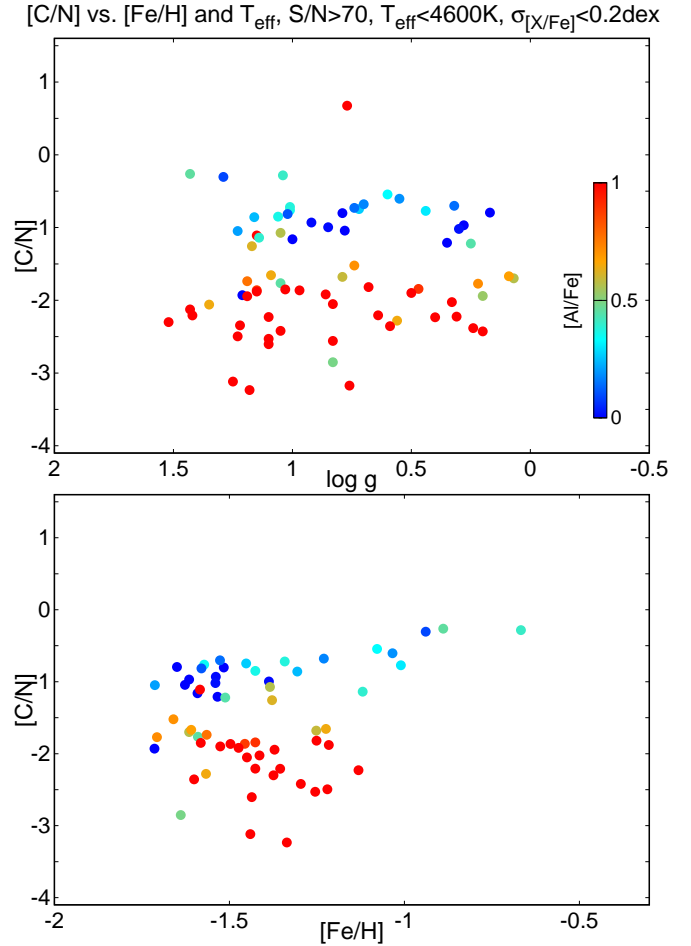}
\caption{[C/N] as a function of metallicity and surface gravity color coded by [Al/Fe] to distinguish between 
FG and SG stars. Correlations in the opposite direction are observed between FG and SG stars as a function of metallicity.
}
\label{fig:cn}
\end{figure}

\begin{figure}
\centering
\includegraphics[width=3.45in,angle=0]{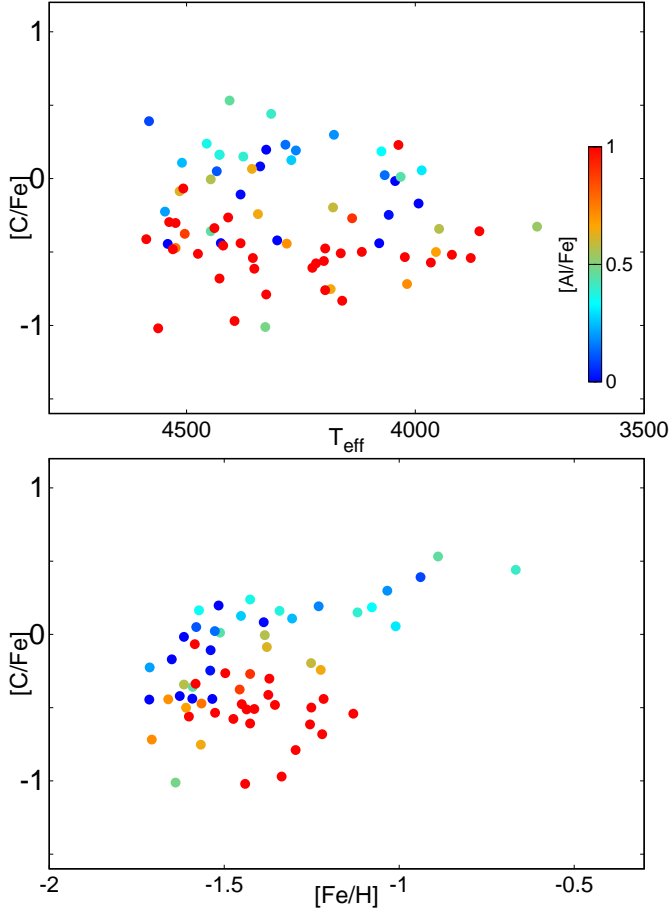}
\caption{[C/Fe] as a function of metallicity and effective temperature color coded by [Al/Fe] to distinguish between 
FG and SG stars. [C/Fe] of FG stars slightly increases with metallicity. 
}
\label{fig:cfeh}
\end{figure}

The slope of the N-C anticorrelation is not only affected by the pollution scenario, but also by the deep mixing 
that occurs on the RGB. The observed slopes discussed above have not been corrected for deep mixing, but \ocen \ 
also shows evidence of mixing as [C/Fe] is strongly correlated with temperature. 

Deep mixing can be investigated by probing the variation of [C/N] abundance ratios as functions of metallicity 
and effective temperature; this is shown in Figure~\ref{fig:cn}.
At first glance the overall value of [C/N] in the studied \ocen \ targets appears to be roughly constant with 
the surface gravity (and also effective temperature) and to a certain degree with metallicity. However, a more complete picture of the 
behavior of [C/N] emerges when stars are divided into the FG and SG stars and also considering the [Al/Fe] 
values (the color bar in Figure~\ref{fig:cn}). Stars with low [Al/Fe] generally have high [C/N] ratios, 
while SG stars with high [Al/Fe] have low [C/N]. This can be understood as the SG stars are formed from 
gas cloud whose composition has been altered by FG stars, increasing the amount of N and resulting in 
lower values of [C/N]. The effects of deep mixing does not appear to be visible as a function of surface 
gravity as [C/N] does not correlate with it, 
but a slight correlation may exist with metallicity when looking at the FG and SG stars separately. 
The [C/N] ratios of FG stars 
increase slightly with increasing metallicity, while the [C/N] of SG stars show the opposite trend, 
decreasing with increasing metallicity. While the [C/Fe] ratio of SG stars is constant with [Fe/H], the 
decreasing trend of [C/N] in SG stars comes from the increased [N/Fe] measured for stars with metallicities 
between [Fe/H] = $-$1.4 and $-$1.2. Thus, the correlation seen when looking at only the SG is not the 
result of deep mixing, but most likely the result of the increased pollution observed at $-$1.4$<$[Fe/H]$<-$1.2.

Model calculations by \citet{lagarde01} suggest that a drop [C/N] should occur after 
the luminosity bump at Teff $\sim$ 4700~K, but this prediction cannot be probed from APOGEE spectra as 
[C/N] is not measurable at such high effective temperatures. However, \ocen \ provides an opportunity 
to investigate the dependence of deep mixing on metallicity independently of mass, since all stars in 
\ocen \ can be considered to have a very similar mass. \citet{lagarde01} also suggested that slightly 
stronger mixing occurs in more metal-rich stars based on theoretical calculations. This would correspond 
to the FG stars that exhibit no extra N from pollution from previous populations. Looking at Figure~\ref{fig:cn}, 
the [C/N] of FG stars indeed  increases slightly with increasing metallicity. While [N/Fe] is constant as a 
function of metallicity, [C/Fe] is not, so this correlation is the result of increased [C/Fe] at high 
metallicities seen in Figure~\ref{fig:cfeh}. Overall, we conclude that 
the study of deep mixing is difficult in \ocen, or in GCs in general, as one has to effectively separate 
multiple populations from the deep mixing.

\subsection{C+N+O}

\begin{figure}
\centering
\includegraphics[width=2.45in,angle=270]{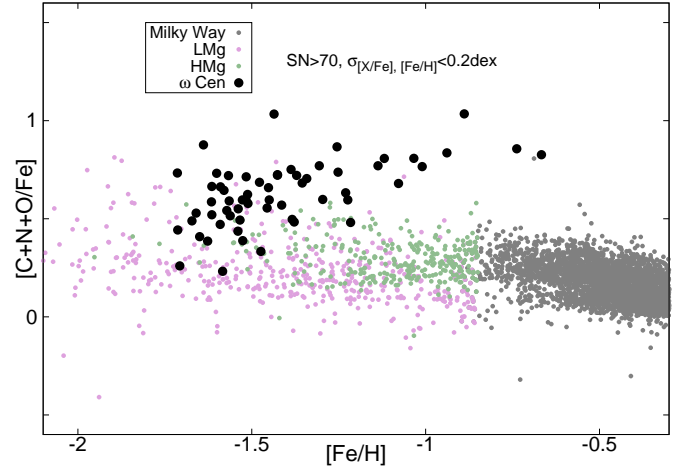}
\caption{The sum of C+N+O in \ocen \ (black dots) and the Milky Way. 
}
\label{fig:cno}
\end{figure}

\begin{figure}
\centering
\includegraphics[width=3.45in,angle=0]{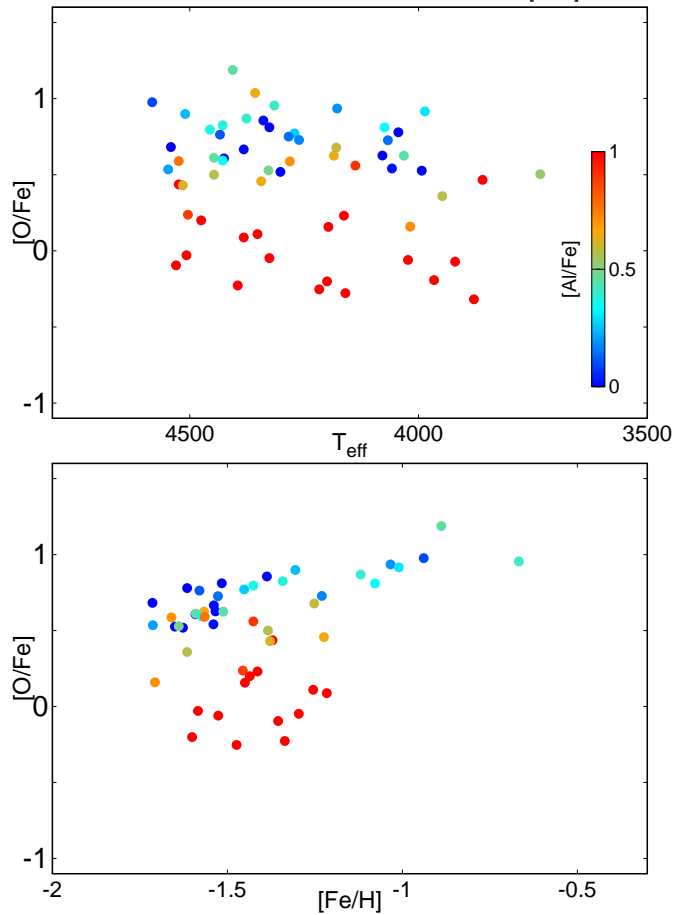}
\caption{[O/Fe] as a function of metallicity and effective temperature color coded by [Al/Fe] to distinguish 
between FG and SG stars. [O/Fe] of FG stars slightly increases with metallicity. 
}
\label{fig:ofeh}
\end{figure}

The C+N+O in \ocen \ has been studied in detail by \citet{marino01}. They found that [(C+N+O)/Fe] correlates 
with metallicity and it increases by nearly 0.5~dex from [Fe/H]=$-$2.0 to $-$0.9. The sum of C+N+O in this 
study is shown in Figure~\ref{fig:cno} as a function of metallicity and also compared to the Milky Way. 
Keeping in mind that the metallicity range in which we can analyze 
the molecular lines in the APOGEE region is more metal-rich than \citet{marino01}, from [Fe/H]=$-$1.7 to 
$-$0.6, we are able to confirm the their findings that the C+N+O correlates with metallicity. This is 
driven mostly by the increased amount of O plotted in Figure~\ref{fig:ofeh} (and to a lesser extent, C) 
at higher metallicities. As there is more C and O in the FG stars, they produce more N 
explaining the increased N abundances at higher metallicities seen in Figure~\ref{fig:ncfe}.

The existence of the enhanced C+N+O is not currently explained, \citet{marino01} could only explain it by 
SNe coming from an arbitrary initial mass function. Most of the stars with very large C+N+O belong to P7, 
the most metal-rich population in \ocen.

\begin{figure*}
\centering
\includegraphics[width=6.2in,angle=0]{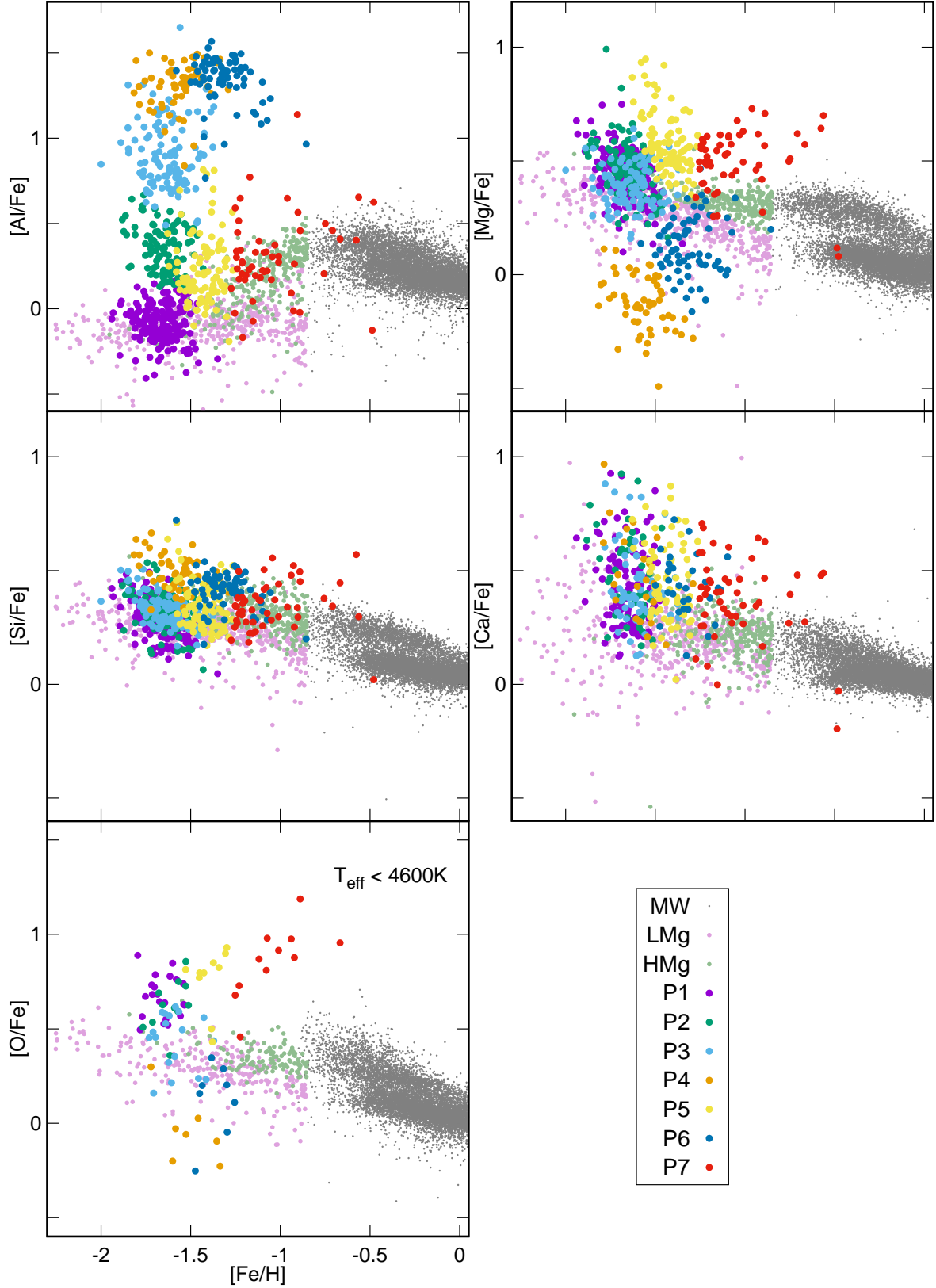}
\caption{The distribution of [Al/Fe], [Mg/Fe], [Si/Fe], [Ca/Fe] and [O/Fe] as a function of [Fe/H] in \ocen \ and the Milky Way. Light green and purple 
dots represent the HMg (metal-poor thick disk) and LMg groups (halo) in our Galaxy as defined by \citet{hayes01}. 
The coloring of the populations in \ocen \ is identical to that of the bottom panel of Figure~\ref{fig:alpop}. Unclassified 
stars are not shown.}
\label{fig:mw1}
\end{figure*}

\section{The Stellar Populations of \ocen \ }

\subsection{An Overview of the Seven Populations}

\begin{figure*}
\centering
\includegraphics[width=6.2in,angle=0]{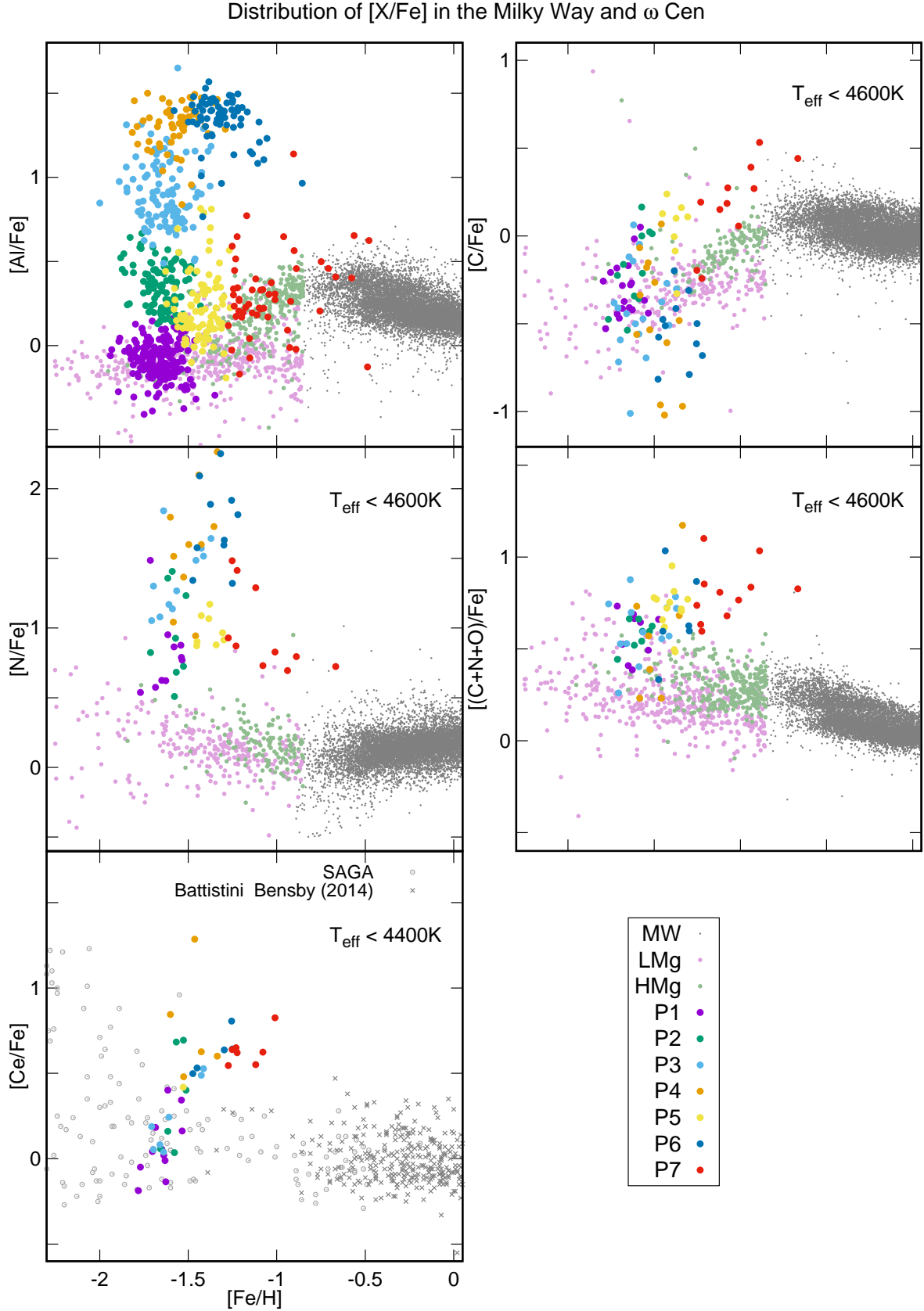}
\caption{The distribution of [Al/Fe], [C/Fe], [N/Fe], [(C+N+O)/Fe] and [Ce/Fe] as a function of [Fe/H] in \ocen \ and the Milky Way. Light green and purple 
dots represent the HMg (metal-poor thick disk) and LMg groups (halo) in our Galaxy as defined by \citet{hayes01}. 
The coloring of the populations in \ocen \ is identical to that of the bottom panel of Figure~\ref{fig:alpop}. Unclassified 
stars are not shown.
}
\label{fig:mw2}
\end{figure*}

A comparison of the chemical abundances in Milky Way populations can now be made with the 
seven populations defined here on the basis of their Al-Fe patterns.  The seven populations (P1 through P7) 
are shown in Figures~\ref{fig:mw1} and~\ref{fig:mw2}, where values of [X/Fe] are plotted versus [Fe/H] for nine elements (Al, Mg, 
Si, Ca, O, C, N, C+N+O, Ce) whose abundances are derived either here or from \citet{meszaros05}. The 
small points in these figures represent the Milky Way populations of thin and thick disks (grey), the 
metal-poor thick disk (green), and halo (pink); the metal-poor thick disk and halo stars are delineated 
according to the criteria from Hayes et al. (2018) of low-Mg (LMg) and high-Mg (HMg) sequences defining 
the halo and metal-poor thick disk populations, respectively. In the top left panel of both figures are 
the values of [Al/Fe] versus [Fe/H], with the seven \ocen \ populations identified by color. 
Closer examination of the populations confirms the visual impression that P1, P2, P3, and P4 are characterized 
by a nearly uniform iron abundance; average values confirm this, with $<$[Fe/H]$>_{\rm P1}=-$1.64$\pm$0.12, 
$<$[Fe/H]$>_{\rm P2}=-$1.60$\pm$0.11, $<$[Fe/H]$>_{\rm P3}=-$1.64$\pm$0.14, $<$[Fe/H]$>_{\rm P4}=-$1.64$\pm$0.11.  The standard 
deviations of these values of [Fe/H] are not much larger than the uncertainties in the iron abundances 
themselves.  In this sense, P1, P2, P3, and P4 resemble closely many typical globular cluster populations, 
which exhibit a large spread in [Al/Fe] abundances, while having only small (if any) [Fe/H] variations. 
Referring back to Figure~\ref{fig:met}, we note that these populations account for $\sim$2/3 of the \ocen \ stars.

The remaining three populations identified here (P5, P6, and P7) are more enigmatic in terms of their possible 
relationship, if any, to the main \ocen \ population (P1 through P4). Looking first at their respective 
Fe abundances reveals that P5 is, within the APOGEE uncertainties, single-valued in iron, with 
$<$[Fe/H]$>_{\rm P5}=-$1.38$\pm$0.04, while P6 and P7 display large ranges in their Fe abundances: 
$<$[Fe/H]$>_{\rm P6}=-$1.17$\pm$0.29 and $<$[Fe/H]$>_{\rm P7}=-$1.05$\pm$0.27. Based on the [Al/Fe] and [Fe/H] 
abundance ratios, P5 and P7 tend to follow the Milky Way halo and thick disk trends, as shown in 
Figure~\ref{fig:mw1}; however their [Mg/Fe] values find them falling well above the Milky Way halo and thick 
disk trends, strengthening the case for \ocen \ being a captured system.  This trend is born 
out by the lower panel of Figure~\ref{fig:mw1}, showing that both P5 and P7 also fall well above the Milky Way 
trends for [O/Fe] versus [Fe/H].  Magnesium and oxygen are produced by, primarily, hydrostatic helium 
($^{16}$O) and carbon ($^{24}$Mg) burning in massive stars and these overabundances, relative to Fe, 
in P5 and P7 may indicate significant enrichment from very massive stars, perhaps as a result of a 
starburst. Unlike P1, P2, P3, and P4, the majority of stars in P5 and P7 do not show evidence of 
having been formed from material exposed to H-burning via the Mg-Al cycle, although both populations 
seem to contain a small fraction of Al-rich stars.  

Compared to the other populations, P6 is unique in being offset to  a higher Fe-abundance, when compared to 
populations P1 through P4, yet composed of stars that were all formed from material that is Al-rich and Mg-poor. 
It may be that P6 is linked to P1, P2, P3, and P4 by the simple addition of Fe from SN (perhaps SN Ia) increasing 
[Fe/H] in P6. 

Other elements can be used to further illuminate and constrain a picture of chemical evolution within \ocen. 
The middle left panel in Figure~\ref{fig:mw1} shows the behavior of [Si/Fe], with all populations in \ocen \ generally 
following the Milky Way halo and thick disk evolution.  Close inspection does reveal the signature of Si 
production in P4 and P6, where these populations fall above the trend of [Si/Fe] with [Fe/H] by $\sim$+0.1 dex. 
Ventura et al. (2016) find that Si production resulting from H-burning in IMS occurs only in metal-poor 
([M/H]$\sim-$3.5) massive AGB stars (M$\sim$5M$_{\odot}$).  The \ocen \ results for [Ca/Fe] are quite 
scattered, likely due to the weakness of the Ca I lines in the APOGEE wavelength window, but follow the Milky Way trends.

Abundances from the heavy s-process element cerium are shown in the bottom panel of Figure~\ref{fig:mw2} and provide 
further information on chemical evolution within the \ocen \ populations.  The populations P1 and P2 
have [Ce/Fe] abundances that scatter around values of $\sim$0.0, while all of the other populations show 
generally enhanced Ce abundances.  Thermally-pulsing AGB (TP-AGB) stars are the main producers of the s-process 
elements, including Ce, pointing to significant contributions from AGB stars to the chemical evolution within \ocen.
Our observation reinforces the findings of \citet{norris02, smith01} who suggested that 
the pollution from low mass AGB stars played a crucial part in the evolution of this cluster.

Taken together, the summary abundance results displayed in Figures~\ref{fig:mw1} and~\ref{fig:mw2} provide insights into understanding 
the stellar populations that constitute the \ocen \ system. The pronounced narrow peak in the [Fe/H] 
distribution (near [Fe/H]$\sim$-1.6) in Figure~\ref{fig:met} is composed of stars that display a strong Al-Mg anticorrelation 
occurring over a very narrow range in Fe-abundance (as well as a N-O anticorrelation and an Al-Si correlation), 
which are signatures that typify many globular clusters.  These stars form the population sequence of P1, P2, P3, 
and P4, as well as P6 (whose stars have significantly larger Fe-abundances and a significant abundance spread): 
these populations contain over 2/3 of the stars within \ocen.  The populations P5 and P7 have different 
chemical signatures and do not contain the strong globular cluster Al-Mg anticorrelation. P5 and P7 both exhibit 
large Fe-abundance spreads, as well as [Mg/Fe] and [O/Fe] abundances that fall well above those of the Milky 
Way halo and thick disk.  Both of these populations also exhibit large [Ce/Fe] abundance ratios, with the 
possibility that P5 and P7 are related stellar systems.

Distilling the abundance patterns in Figures~\ref{fig:mw1} and~\ref{fig:mw2}, into, perhaps, the simplest scenario to explain the 
seven populations defined by the [Al/Fe] abundance ratios would be that \ocen \ consists of two general 
populations.  The first population (consisting of P1, P2, P3, P4, and P6) is the underlying globular cluster 
system that defines the classification of \ocen \ as a globular cluster, but one that is very massive 
and contains a significant spread of Fe-abundances (driven by P6).  In addition, there is a second distinct 
population (P5 and P7) characterized by large [hydrostatic-$\alpha$/Fe] abundances, as well as large s-process 
abundances.  Chemical evolution models would require that P5 and P7 contain material that has been cycled 
through very massive stars (perhaps a star burst) mixed with ejecta from low-mass TP-AGB stars, which evolve 
over at least a few Gyr.  These two intermingled populations, P5 and P7, within \ocen \ presumably 
represent the central nuclear remnants of a disrupted and captured dwarf galaxy. Such a comparison cannot 
be made explicitly for \ocen, as we do not observe the parent dwarf galaxy, but a clear connection between 
the chemical properties of M54 and the Sagittarius dwarf spheroidal galaxy was explored before showing that 
interaction between GCs and dwarf galaxies happened in the past \citep{mucciarelli01}.

\subsection{Comparing with the Milky Way}

The discussion in the previous section sketched out possible relationships between the 
seven \ocen \ populations identified here based on their various elemental abundance patterns, a process known 
as chemical tagging. A comparison of chemical abundances in the \ocen \ populations with those of the Milky Way, 
via chemical tagging, is a useful exercise to illuminate both similarities, as well as differences, in chemical 
evolution within the two stellar systems.
Chemical tagging uses the idea that detailed abundance measurements can be used to identify spatially separated 
stars that were born in the same molecular cloud, as first presented by \citet{hogg01} using APOGEE data. 
For our chemical tagging, stars from the Milky Way were selected by applying the criteria defined by 
\citet{hayes01} to the DR16 data \citep{ahumada01}. \citet{hayes01} has divided the metal-poor stars into
two groups based on the Mg abundances that are separated from each other by a low density gap. At low metallicities,
[Fe/H]$< -$0.9, the [Mg/Fe]=$-$0.2$\cdot$[Fe/H] was used by \citet{hayes01} to separate the low-[Mg/Fe]
population from the high-[Mg/Fe] one. Since we have updated the data to DR16, which has a
slightly different abundance scale, this selection function was also updated to [Fe/H]$< -$0.85 to still avoid most
of the thick disk stars, and [Mg/Fe]=$-$0.29$\cdot$[Fe/H]$-$0.07 to account for the new position of the gap, which starts
to open up in the [Fe/H]=$-$1.4 to $-$1.3 range. 

It is important to note that while both our analysis and DR16 uses APOGEE data, the abundances were derived independently 
from each other. Our analysis method is described in detail by \citet{masseron01}, and the differences amount to about 
an average 0.1 dex shift in metallicity and 0.2 dex shift in [Al/Fe] between BACCHUS and ASPCAP DR16. These 
discrepancies also depend on effective temperature, because of differences between photometric and ASPCAP 
temperatures. We do not calibrate our results to those of ASPCAP, but when doing comparisons between \ocen \ 
and the Milky Way one must keep in mind that such small discrepancies in abundances exist.

Inspection of the various elemental abundances in Figures~\ref{fig:mw1} and~\ref{fig:mw2} reveals that \ocen \ is quite different in 
the behavior of most elements as a function of Fe-abundance.  The large numbers of second generation (SG) stars 
in P2, P3, P4, and P6 reveal strong signatures of hot H-burning in their Mg and Al abundances, as well as in 
their N and O abundances (as do most Galactic globular clusters).  Even the populations dominated by first 
generation (FG) stars (P1, P5, and P7) exhibit different behaviors, relative to the Milky Way, in their Mg, Al, 
N, and O abundances, with these values, relative to Fe, tending to fall above the Milky Way trends.  The larger 
values of [O/Fe] and [Mg/Fe] are likely due to larger contributions from massive stars, via SN II, relative to 
the Milky Way, as noted in the previous section.  The larger values of [N/Fe], [C/Fe] point to significant material 
that has been cycled through intermediate-mass stars (IMS), with the additional steep increase in [Ce/Fe] created by 
these same IMS stars, along with lower-masses, evolving through the TP-AGB phase of stellar evolution.
It is only [Si/Fe] that behaves similarly to the Milky Way, excluding the extreme SG populations of P4 and P6, which 
exhibit Si enhancement from hot H-burning.  The results for [Ca/Fe] are more scattered, making a definitive comparison 
difficult.

The comparison of the detailed chemical abundance patterns within the numerous \ocen \ populations with those of the 
Milky Way (chemical tagging) reveals a unique mixture of chemical evolutionary signatures, peculiar to \ocen.  
There is evidence of significant chemical evolution due to SN II, with perhaps larger contributions from the more 
massive stars (with larger values of O and Mg, relative to Si), while at the same time, there is evidence for 
substantial chemical evolution dominated by low and intermediate mass AGB stars.  Adding to the complexity and 
peculiarity of \ocen \ is the mixture of both globular cluster abundance signatures (P1, P2, P3, P4, and P6) 
along with P5 and P7, which do not show such extreme Al, Mg, and N abundance variations.

\begin{figure*}
\centering
\includegraphics[width=6.2in,angle=0]{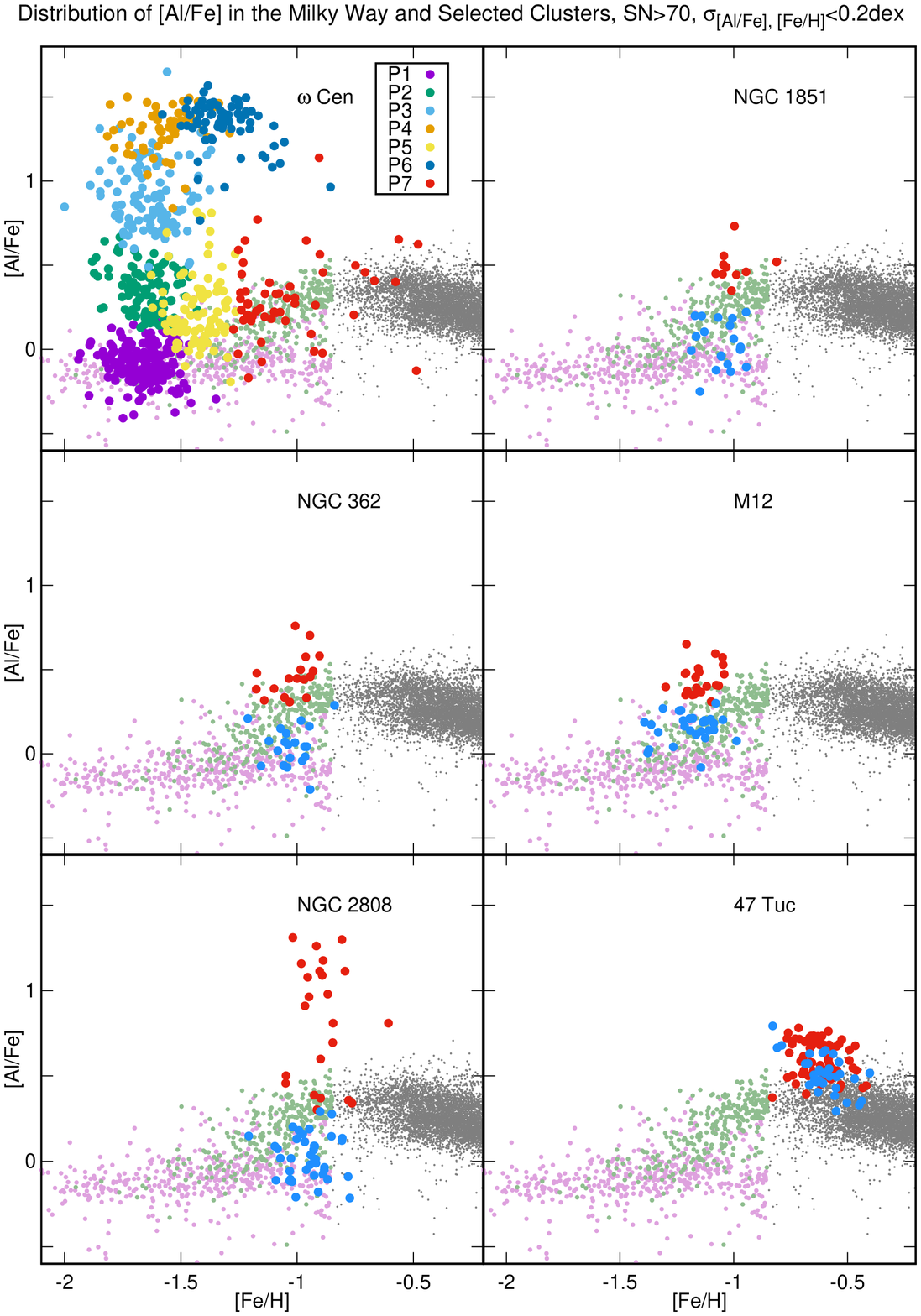}
\caption{The distribution of [Al/Fe] as a function of [Fe/H] in NGC~1851, NGC~362, NGC~2808, M5 and 47~Tuc and the Milky Way. 
Green and purple dots represent the HMg (metal-poor thick disk) and LMg groups (halo) in our Galaxy as defined by \citet{hayes01}. 
Blue symbols denote FG stars with [Al/Fe]$<$0.3, red dots are used for SG stars  with [Al/Fe]$>$0.3. Because Al is nearly 
constant in 47~Tuc we use [N/Fe]$>$0.7 to separate SG stars from the FG ones.
}
\label{fig:mw_al}
\end{figure*}

\subsection{Comparison with Other Globular Clusters}

It is of interest to compare the \ocen \ results for [Al/Fe] as a function of [Fe/H] with those from other globular clusters 
of the Milky Way. This is shown in Figure~\ref{fig:mw_al} for the globular clusters NGC 1851, NGC 362, M 12, NGC 2808 and 47 
Tuc. The GCs NGC~1851 and NGC~362 were selected because these show Ce abundance enhancements similar to \ocen \ 
\citep{meszaros05}, indicating that there has been pollution from low mass AGB stars in these clusters. M12 was selected 
because it covers the metallicity range where the Low-Mg and High-Mg groups segregate, although this cluster shows only a 
small spread in the Al abundances; NGC~2808 covers the metallicity range ($-$1.1$<$[Fe/H]$<-$0.8) where \ocen \ has no, 
or very few stars with halo-like [Al/Fe] values, and 47~Tuc is an example of a more metal-rich globular cluster with no 
significant Al abundance scatter. There are other clusters with [Fe/H]$>-$1.6, for example M107 and M71, but the five 
selected GCs are representative of the Al abundance distributions found in other GCs. For a more detailed discussion on 
other clusters we refer to \citet{meszaros05}.

Most of the FG stars of NGC~1851 and NGC~362 have [Al/Fe] values that are similar to the halo, while M12 is interesting 
as it shows similarity to \ocen \ in that almost 
all of its FG stars have larger, thick disk-like [Al/Fe] ratios.
But this similarity ends here as M12 does not have Ce enhancement, thus no low-mass AGB pollution unlike NGC1851, NGC362, 
and \ocen. 47~Tuc formed from a thick disk like cloud with no obvious further 
Al enrichment visible, the scatter is on the level of the average uncertainty of [Al/Fe]. Thus, our conclusion is that 
M12 and 47~Tuc formed from interstellar gas that had a similar Al/Fe composition to the thick disk.

None of the five clusters show a similar distribution to \ocen, but these observations show that NGC~1851, NGC~362
and NGC~2808 started with halo-like FG stars and internally self-enriched SG stars with no additional gas added from 
the thick disk. 
We believe this is the case because there are no clear Al peaks in their Al histogram at around [Al/Fe]=0.3$-$0.4, 
where the thick disk concentrates. NGC~2808 have a small peak at [Al/Fe]=0.39 \citep{meszaros05}, but that is based 
on a sample of only six stars. 
Note that \citet{carretta08} did find a population of 28 
stars centering around [Al/Fe]=0.206, however most of those stars lie between the thick disk and halo in Figure~\ref{fig:mw_al} 
in our analysis.


\begin{figure}
\centering
\includegraphics[width=3.45in,angle=0]{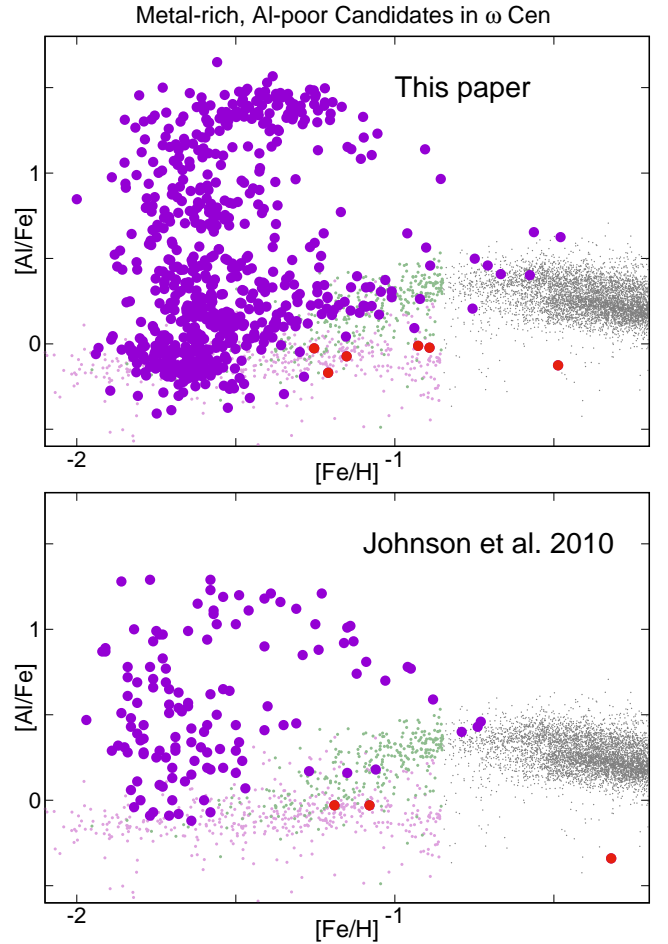}
\caption{Metal-rich stars with low [Al/Fe] values are denoted by red dots from both our study (top panel) and that of 
\citet{johnson02} (bottom panel). There are no common stars in the two studies, thus their existence must be confirmed 
by subsequent observations.
}
\label{fig:alcomp}
\end{figure}

Overall, we can conclude that none of the APOGEE sampled GCs have similar abundance patterns to that of the P7 group in 
\ocen: we do not find any existing GCs that replicate the properties of the most metal-rich population in \ocen.


\subsection{The Case of P7: Interaction with the Thick Disk?}

As discussed in the previous three sections, the most metal-rich population classified in this study, P7, presents a 
challenge in establishing its relationship to the other populations.  The possibility that \ocen \ is composed of a 
collection of captured, or merged components is supported by the observations of \citet{jurcsik01, pancino02, 
pancino03, calamida01}, all of whom showed that the more metal-rich stars are spatially different from the metal-poor 
ones, including stars from P7. While there are some discrepancies between these literature sources about the exact 
details in the accurate structure of the spatial coverage, all of them agree that the metal-rich stars with 
[Fe/H]$>-$1.2 (RGB-Int2+3 and RGB-a groups) are more concentrated in the southern part of the cluster. We note 
that the discrepancy of spatial coverage of the different metallicity groups is not visible from our data alone, 
but our sample (982 stars) is not only orders of magnitudes smaller than the photometric surveys mentioned above, 
but also does not contain any stars from the inner 10$-$15 arcmin part of the cluster. The connection between the 
spatial coverage and our observation of how the Al behaves in the FG stars in these 
metal-rich groups may suggest the idea that the majority of the metal-rich FG stars formed from gas originating 
from an environment distinct from that of the more metal-poor populations.  

In earlier discussion here, we noted that P7 resembles the thick disk in the values of [Al/Fe] over the Fe-abundance 
range spanned by P7; however, when other elemental ratios are considered, such as [O/Fe], [Mg/Fe], [C/Fe], or 
[Ce/Fe], the trends defined by this suite of elements all fall well above the Milky Way thick disk. This 
suggests that P7 has no chemical connection to a Milky Way population, and the merging, or capture with the 
dominant populations of \ocen \ occurred before the assimilation of \ocen \ into the inner reaches of the Milky Way.

There remains an additional peculiarity about P7, in that it contains six stars (2M13262643-4736438, 
2M13275189-4750038, 2M13272565-4715345, 2M13253337-4730395, 2M13262451-4709113, 2M13260181-4739340 ) 
that have low values of [Al/Fe], close to that of the halo and considerably lower than the rest of the 
P7 stars. These stars may indicate a more complex origin for the metal-rich component of \ocen. 
But, we must be careful in assessing if these stars are indeed metal-rich 
and Al-poor. After checking our abundance analysis we identified four stars that have unusually 
high photometric temperatures compared to ASPCAP (T$_{\rm eff} >$500K). If their photometric 
temperatures turned out to too high due to photometric errors, then their 
metallicities would be lower and those stars would not pertain any more to the peculiar 
metal-rich and Al-poor group. Yet, this still leaves us with two stars, 2M13275189-4750038 
and 2M13272565-4715345, the two most metal-rich out of these six, for which the temperature 
differences are smaller, and thus their metallicities more reliable. \citet{johnson02} 
have also observed three stars (50187, 55028, 60058) that satisfy our criteria to be metal-rich and al-poor. 
Unfortunately, we do not have any of these stars in common between the two studies, thus we cannot 
reliably conclude that either the observations of \citet{johnson02}, or ours have found halo-like 
stars in the most metal-rich population of \ocen. Nevertheless, the fact the two independent 
studies found stars like these hints at their existence and more observations are needed to 
firmly confirm their abundances.


\section{Summary}

We investigated the multiple populations of \ocen \ by analyzing the observational effects of the MgAl and CNO cycles 
using APOGEE data, 
Based on our findings we conclude the following:

1. The four metallicity groups found by \citet{johnson02} are confirmed, and we found seven populations based on their 
[Fe/H], [Al/Fe] and [Mg/Fe] abundances. This confirms the findings of \citet{gratton04} by using different elements to 
trace populations.

2. We find that the shape of the Al-Mg anticorrelation clearly depends on metallicity, the metal-poor groups ([Fe/H]$<-$1.2) show 
continuous, the metal-rich groups ([Fe/H]$>-$1.2) bimodal distributions. 

3. We find evidence of that the evolution of Al in the FG stars is very similar to that of thick disk of the Milky Way 
by comparing the [Al/Fe]-[Fe/H] distribution of \ocen \ with that of our Galaxy, but our findings of elevated $\alpha-$elements, 
CNO, and Ce suggests that P7 has no chemical connection to a Milky Way population, and the merging, or capture with the 
dominant populations of \ocen \ occurred before the assimilation of \ocen \ into the inner reaches of the Milky Way.

4. There are six stars observed in P7 that have [Al/Fe]$<$0.0. These may suggest that \ocen \ could be the remnant core of 
a larger dwarf galaxy, but the existence of these stars needs an independent confirmation.

5. We report that the N-C anticorrelation also depends on metallicity, similarly to the Al-Mg anticorrelation. The 
distribution is continuous up to [Fe/H]$<-$1.2, then becomes bimodal at higher metallicities. 

6. The two populations have different [C/N] values. We may observe a slight positive correlation between [C/N] and metallicity 
in the FG stars. 

7. The increased C+N+O with increased metallicity previously found in the literature is confirmed. 

\section{Data availability}

The data underlying this article are available in the article and in its online supplementary material.

\acknowledgements{SzM has been supported by the J{\'a}nos Bolyai Research Scholarship of the Hungarian Academy of
Sciences, by the Hungarian NKFI Grants K-119517 and GINOP-2.3.2-15-2016-00003 of the Hungarian National
Research, Development and Innovation Office, and by the {\'U}NKP-19-4 and {\'U}NKP-20-5 New National Excellence Program of the 
Ministry for Innovation and Technology from the source of the National Research, Development and Innovation Fund. 
DAGH, and TM acknowledge support from the State Research Agency (AEI) 
of the Spanish Ministry of Science, Innovation and 
Universities (MCIU) and the European Regional Development Fund (FEDER) under grant AYA2017-88254-P. 
J.G.F-T is supported by FONDECYT No. 3180210. 
D.G. gratefully acknowledges support from the 
Chilean Centro de Excelencia en Astrof{\'{\i}}sicay Tecnolog{\'{\i}}as Afines (CATA) BASAL grant AFB-170002. D.G. also acknowledges financial support from the Direcci\'on de Investigaci\'on y Desarrollo de
la Universidad de La Serena through the Programa de Incentivo a la Investigaci\'on de
Acad\'emicos (PIA-DIDULS). T.C.B. acknowledges partial support from grant PHY 14-30152, Physics Frontier Center/JINA Center for the
Evolution of the Elements (JINA-CEE), awarded by the US National Science Foundation.

Funding for the Sloan Digital Sky Survey IV has been provided by the Alfred P. Sloan Foundation, 
the U.S. Department of Energy Office of Science, and the Participating Institutions. SDSS-IV acknowledges
support and resources from the Center for High-Performance Computing at
the University of Utah. The SDSS web site is www.sdss.org.

SDSS-IV is managed by the Astrophysical Research Consortium for the 
Participating Institutions of the SDSS Collaboration including the 
Brazilian Participation Group, the Carnegie Institution for Science, 
Carnegie Mellon University, the Chilean Participation Group, the French Participation Group, 
Harvard-Smithsonian Center for Astrophysics, 
Instituto de Astrof\'isica de Canarias, The Johns Hopkins University, 
Kavli Institute for the Physics and Mathematics of the Universe (IPMU) / 
University of Tokyo, Lawrence Berkeley National Laboratory, 
Leibniz Institut f\"ur Astrophysik Potsdam (AIP),  
Max-Planck-Institut f\"ur Astronomie (MPIA Heidelberg), 
Max-Planck-Institut f\"ur Astrophysik (MPA Garching), 
Max-Planck-Institut f\"ur Extraterrestrische Physik (MPE), 
National Astronomical Observatories of China, New Mexico State University, 
New York University, University of Notre Dame, 
Observat\'ario Nacional / MCTI, The Ohio State University, 
Pennsylvania State University, Shanghai Astronomical Observatory, 
United Kingdom Participation Group,
Universidad Nacional Aut\'onoma de M\'exico, University of Arizona, 
University of Colorado Boulder, University of Oxford, University of Portsmouth, 
University of Utah, University of Virginia, University of Washington, University of Wisconsin, 
Vanderbilt University, and Yale University.
}


\end{document}